\documentclass[a4paper]{article}
\usepackage[margin=25mm]{geometry}
\usepackage{amsfonts}
\usepackage{amssymb}
\usepackage{graphicx}
\usepackage{svg}
\usepackage{amsmath}
\usepackage{caption}
\usepackage{etoolbox}
\usepackage{bm}
\usepackage[labelfont={small,bf}]{caption}
\usepackage[font={small}]{caption}
\usepackage{subcaption}
\usepackage{hyperref}
\usepackage{authblk}
\usepackage[numbers,sort&compress]{natbib}

\title{Robust chemotaxis beyond sensing limits: signal, noise, and strategy}

\author{Robert G. Endres${}^\dagger$}

\affil{Department of Life Sciences, Imperial College, London SW7 2AZ, United Kingdom\newline\newline${}^\dagger$ Email: r.endres@imperial.ac.uk}

\begin{document}
\maketitle

\begin{abstract}
Bacterial chemotaxis has long been viewed as operating near the physical limits of sensing, as originally articulated by Berg and Purcell. Recent information-theoretic analyses challenge this view, suggesting that {\it Escherichia coli} uses only a small fraction of the information available in ligand arrival statistics to bias its motion. How should such low information efficiency be interpreted at the level of behavior? Here, I argue that chemotactic performance is shaped not only by information transmission and noise, but by the strategy of movement itself. Using simple scaling arguments and minimal models, I show how run-and-tumble chemotaxis can remain robust to noise through symmetry and temporal averaging, even when internal information processing is inefficient. Comparing bacterial and eukaryotic chemotaxis highlights how different sensing strategies convert physical limits into observable behavior. These considerations suggest that low information efficiency need not imply poor performance, but may instead reflect an evolved balance between robustness, simplicity, and function.
\end{abstract}

Bacteria navigate chemical gradients by relying on noisy molecular cues, using a biased random walk based on run-and-tumble motion \cite{Berg2000MotileBehavior,SourjikWingreen2012Chemotaxis}. While larger eukaryotic cells perform direct spatial gradient sensing across their cell body, small fast-swimming bacteria perform temporal gradient sensing, that is, by comparing concentration measurements in time \cite{lovely1975jtheorbio,alonso2025pnas,rode2024prxlife}. Classical theory has long suggested that this process of chemotaxis operates close to fundamental physical limits of sensing, first articulated for the bacterium {\it Escherichia coli} by Berg and Purcell in 1977 \cite{berg1977biophysj}. These limits arise from the stochastic arrival of ligand molecules at the cell surface by diffusion, an unavoidable source of noise that sets a baseline uncertainty for any measurement (see \cite{AquinoWingreenEndres2016ThinkAgain} for a review). Additional fluctuations can be introduced downstream by biochemical signaling networks, which often operate with small numbers of molecules \cite{BialekSetayeshgar2005Limits}. 

Berg and Purcell argued that when framed in terms of signal-to-noise ratios, reliable gradient sensing requires that concentration differences along a "run" exceed the uncertainty associated with these noise sources \cite{berg1977biophysj}. For physiological parameters and shallow gradients, their analysis indeed suggested that bacterial chemotaxis approaches the predicted limit (see also \cite{Brumley2019PNAS} for ocean bacteria). Physical sensing limits are not only relevant for bacteria, but also constrain gradient sensing in larger eukaryotic cells, including amoeba, immune cells, and migrating cancer cells \cite{endres2008pnas}.  However, it remains unclear how such physical limits on sensing accuracy translate into effective chemotactic behavior and efficiency over longer time scales, where persistence, memory, and movement strategy play a central role.

Recently, Mattingly and colleagues revisited this classic problem, applying an information-theoretic framework to quantify how efficiently {\it E. coli} converts stochastic ligand arrivals into directed motion \cite{mattingly2024natphys_submitted}. Their striking result—that cells use only about one percent of the available information—challenges the long-standing view that bacterial chemotaxis operates near the physical sensing limit. This conclusion is particularly surprising because earlier work using closely related formalisms showed that chemotactic performance is information-limited, with reduced information transmission leading to substantially impaired navigation \cite{mattingly2021natphys}. Although phenotypic variability can enhance chemotactic ability in a small subset of cells, the overall finding that typical cells operate far below the information optimum remains unexpected and difficult to interpret.

Here, I argue that chemotactic performance is shaped not only by information transmission and noise, but by movement strategy itself. Using simple scaling arguments and minimal models, I show that the run-and-tumble strategy can cancel much of the external and internal noise through symmetry and temporal averaging, even when internal information processing is inefficient. Comparing bacterial and eukaryotic chemotaxis highlights how different sensing strategies convert physical sensing limits into observable behavior. This analysis places the findings of Mattingly {\it et al.} in a broader biophysical context. More generally, it suggests that low information efficiency need not imply poor performance, but may instead reflect an evolved balance between robustness, simplicity, and function, \textcolor{black}{where simplicity refers to a minimal control strategy in which cells do not explicitly estimate gradient direction.}

\section*{Information-theoretic limits of chemotactic drift}

I begin with a brief recap of the information-theoretic framework introduced by Mattingly {\it et al.} \cite{mattingly2024natphys_submitted}.
In particular, this framework includes causal versions of the mutual information, measuring correlations between inputs and outputs, as well as rate-distortion theory, which quantifies the minimal information required to achieve biased motion up a gradient for a given acceptable noise level. In particular, they quantify how efficiently \emph{E. coli} converts external ligand information given by $s(t)=d/dt\, \log c=g v_x(t)$ into internal signaling activity, with $g=\nabla c/c$ the relative gradient and $v_x(t)$ the effective instantaneous cell swim speed along the gradient in the $x$ direction. 

The paper defines two dimensionless signal-to-noise ratios (SNRs): one for ligand arrival at the cell surface (\( \gamma_r\)) and another for kinase activity response (\( \gamma_a \)). These are used to compute two corresponding steady-state information rates, \( \dot{I}_{\dot{s} \to r}^* \) and \( \dot{I}_{\dot{s} \to a}^* \), capturing the predictive information from ligand arrival to receptors, and from ligand arrival to downstream biochemical signaling, respectively, using the known physical limit of temporal gradient sensing \cite{mora2010prl}. 
The ratio \( \eta = \dot{I}_{\dot{s} \to a}^* / \dot{I}_{\dot{s} \to r}^* \) quantifies the efficiency of information transmission. However, the measurable quantity is the cells' drift velocity up the gradient.

To ease the discussion and extract intuition, I next introduce a simple information-theoretic toy model connecting ligand arrival statistics (input) to drift behavior (output). Note, this toy model is not intended as a replacement for detailed signaling models, but as a bridge between physical limits and measurable behavior. On the one hand, this provides intuition behind the rigorous and long mathematical derivations in Mattingly {\it et al.}'s papers  \cite{mattingly2024natphys_submitted,mattingly2021natphys}. On the other hand, I would like to make cell-behavioral predictions which can be compared with experimental data. In particular, I am interested in scaling relations. 

At the most basic level, information about the temporal stimulus arises from estimating changes in concentration during individual runs with speed $v$. As the cell needs to estimate the stimulus $s(t)$ over a time scale $\tau$ of a run, this results in a concentration change of $\Delta\log(c)=gv\tau$. The noise in this estimate arises from Poisson fluctuations in ligand arrival, with variance $\langle(\delta\log(c))^2\rangle\sim 1/(r_0 \tau)$, where $r_0 \sim c$ is the average arrival rate. Hence, at the input level, the simplified information content about $s(t)$ is \cite{cover2006elements}

\begin{equation}
    I_\text{arrival}\sim \frac{1}{2}\log(1+\text{SNR})\sim\frac{\text{SNR}}{2\ln 2},
\end{equation}
with $\ln 2$ due to conversion from nats to bits. Thus, the signal-to-noise ratio for estimating $\Delta \log c$ is:
\begin{equation}
\text{SNR} \sim \frac{(g v \tau)^2}{1/(r_0 \tau)} = g^2 v^2 r_0 \tau^3.
\end{equation}
Since this estimate is made once every $\tau$ seconds, the information rate becomes:
\begin{equation}
\dot{I}_{\text{arrival}} \approx \frac{\text{SNR}}{2 \ln 2 \cdot \tau} = \frac{g^2 v^2 r_0 \tau^2}{2 \ln 2}.
\end{equation}
%\textcolor{black}{Even though the diffusion-limited capture of ligand molecules can be large based on flux $J \sim 4\pi D a c \sim r_0$ \cite{berg1977biophysj}, implying relatively small fluctuations in absolute concentration measurements, the signal-to-noise ratio  can remain small in shallow gradients. [The present results show that, in this regime, the run-and-tumble strategy can nevertheless render chemotactic drift robust to such fluctuations.]}
\textcolor{black}{Even though the diffusion-limited capture of ligand molecules can be large based on flux $J \sim 4\pi D a c \sim r_0$ \cite{berg1977biophysj}, implying relatively small fluctuations in absolute concentration measurements, the relevant signal-to-noise ratio for temporal gradient sensing behaves as $\mathrm{SNR} \sim g^2 r_0$, which can remain small in shallow gradients.}

At the output level, we quantify the information in behavior. If the cell does simple random walk e.g. by runs and tumbles, $\langle x^2\rangle \sim 2D_\text{eff}t$ with $D_\text{eff}\sim v^2\tau$. In contrast, if there is a bias, the ballistic motion is $\langle x(t)\rangle\sim v_\text{drift} t$ with the mean drift velocity $v_\text{drift}$. To achieve this bias in movement, the cell must reduce the uncertainty of its position over time by some amount, and this reduction requires information. Using rate-distortion theory, one obtains the lower bound is $v^2_\text{drift} \leq 2D_\text{eff} \dot I_\text{behavior}$ \cite{mattingly2021natphys}, from which we obtain

\begin{equation}
    \dot I_\text{behavior}\geq \frac{v^2_\text{drift}}{2D_\text{eff}}.
\end{equation}
Combining the two information rates yields the information transmission efficiency
$\eta = \dot{I}_{\text{behavior}}/{\dot{I}_{\text{arrival}}}$ which produces values within reasonable agreement with the $\sim 1\%$ bound reported by Mattingly {\it et al.} for typical parameters \cite{mattingly2024natphys_submitted}. Hence, this is a meaningful and simple model we can use here.

The existence of a regime in which chemotactic drift becomes independent of input noise motivates a reinterpretation of low information efficiency. As the arrival rate scales with concentration, $r_0 \propto c$, we conclude $\dot{I}_{\text{arrival}} \propto g^2 c$. Assuming a fixed fraction of this information is transmitted behaviorally (i.e., $\dot{I}_{\text{behavior}} \propto \dot{I}_{\text{arrival}}$), we have $\dot{I}_{\text{behavior}} \sim g^2 c$, which leads to
\begin{equation}
v_{\text{drift}} \sim \sqrt{\dot{I}_{\text{behavior}}} \sim g \sqrt{c},\label{eq:v_drift_1}
\end{equation}
and hence $v_{\text{drift}} \sim \sqrt{\text{SNR}}$.
This agrees with the scaling of the full model in Mattingly {\it et al.} for $\gamma_a\rightarrow \infty$ when internal noise is not limiting. In contrast, if internal noise is limiting for $\gamma_a\rightarrow 0$, their comprehensive model instead leads to a slightly different scaling $v_{\text{drift}} \sim g c$ in the  linear sensing regime and $v_{\text{drift}} \sim g$ in the important log-sensing regime. However, the conceptual significance of the latter result was not emphasized: in the log-sensing regime, the drift becomes independent of the SNR, and thus of ligand input noise. This suggests two possible interpretations: either internal noise dominates to the extent that external noise becomes irrelevant, or that cells are remarkably robust to noise. The latter interpretation appears more consistent with the results presented here.

\begin{figure}[t]
\centering
\includegraphics[scale=0.5]{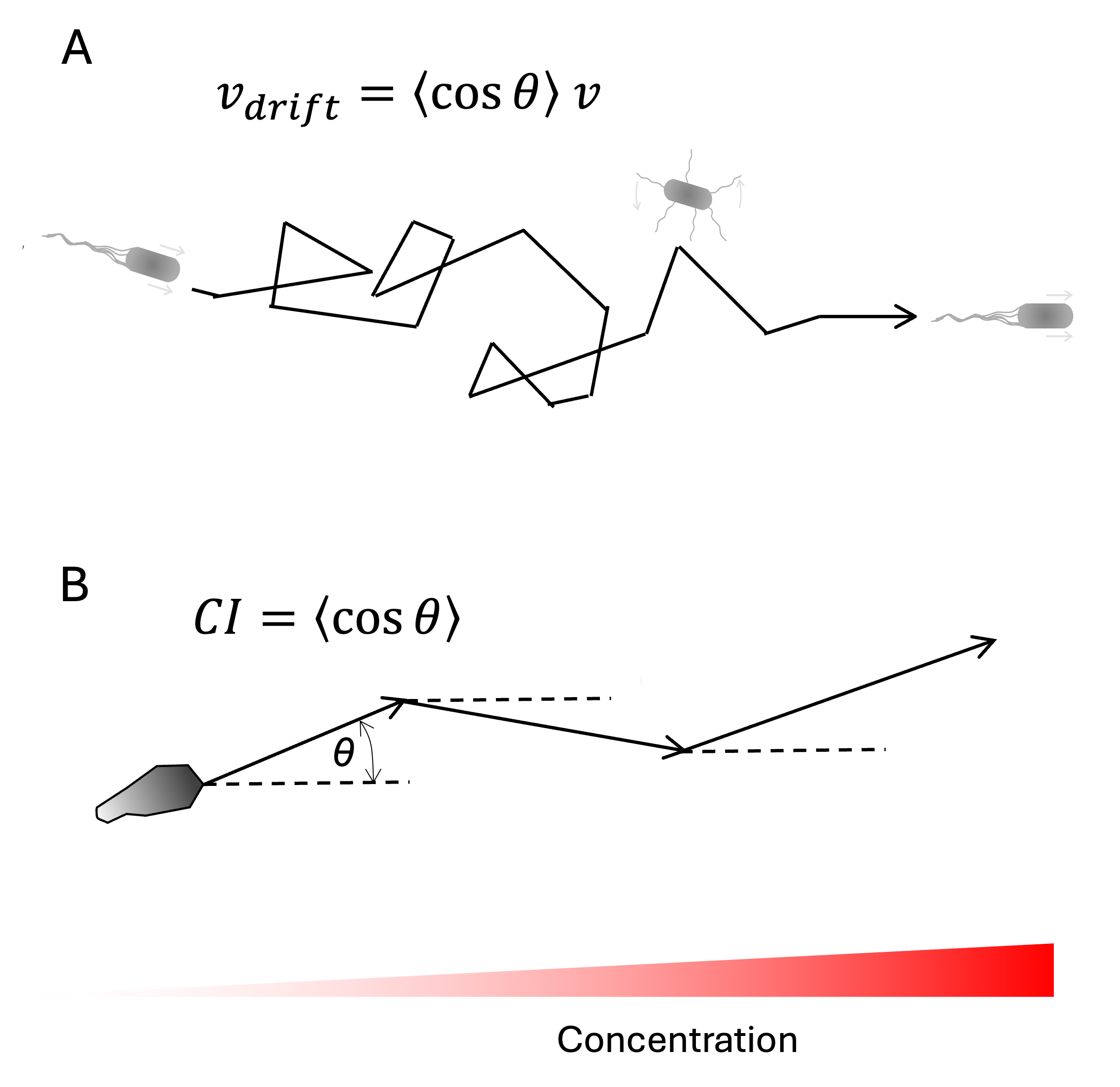}\caption{{\bf Bacterial vs eukaryotic chemotaxis.} 
\textcolor{black}{(A) Drift velocity in bacterial chemotaxis, described by biased random walk up a chemical gradients based on run-and-tumble motion. (B) Chemotactic index in eukaryotic chemotaxis with $\theta$ the direction of movement relative to the gradient. Note drift and CI are closely related up to speed factor $v$.}}
\label{fig:fig1}
\end{figure}

\section*{Comparison to previous mean-field models and Weber's law}

\begin{figure}[t]
\centering
\includegraphics[scale=0.4]{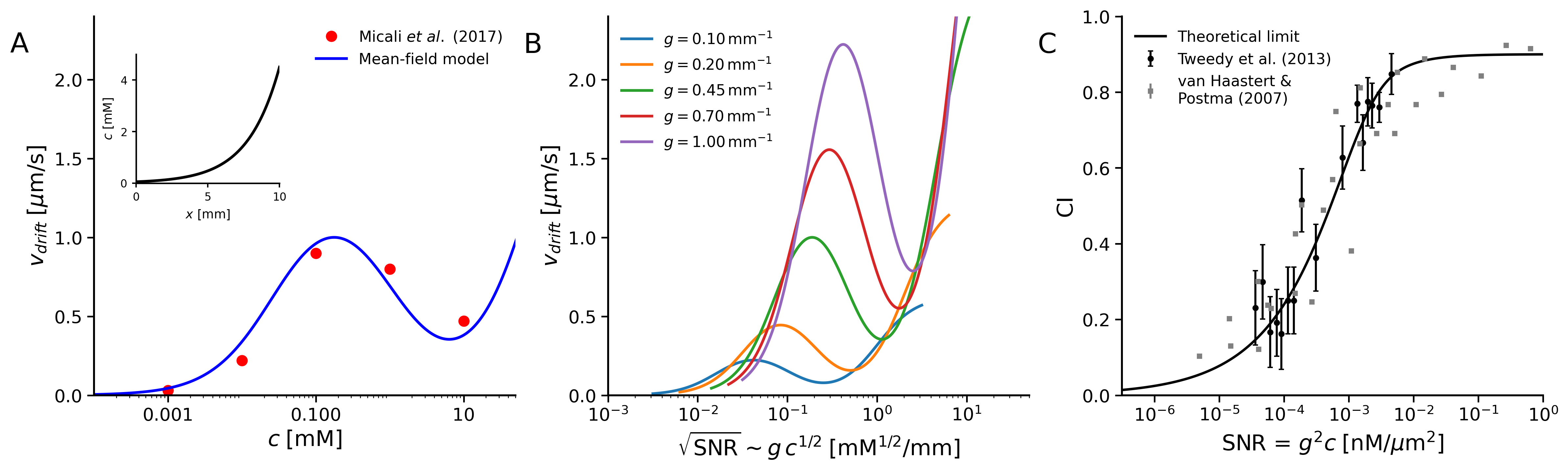}\caption{{\bf Bacterial and eukaryotic chemotaxis compared to data.} 
(A) Experimental data and \textcolor{black}{mean-field model for the drift velocity} in {\it E. coli} chemotaxis for fixed relative gradient $g$ (redrawn from \cite{micali2017biophysj}). The two peaks correspond to the Tar (left) and Tsr (right) receptor types responding to the chemoattractant MeAsp, \textcolor{black}{constituting the logarithmic sensing regimes. Model parameters: $C=9.41$ [mm $\mu$m/s] and others were taken from Ref. \cite{micali2017biophysj}.} The inset shows the corresponding exponential gradient. 
\textcolor{black}{(B) Mean-field model for different relative gradients $g$ and plotted as a function of $\sqrt{\text{SNR}}$. The curves do not collapse except below each peak when $c$ dependence is weak.}
(C) Experimental data for {\it Dictyostelium discoideum} chemotaxing toward cAMP, \textcolor{black}{showing a data collapse when scaled by the SNR}  \cite{van_haastert_biased_2007,tweedy_distinct_2013}.}
\label{fig:fig2}
\end{figure}

To put these results into context, it is useful to compare them with earlier theoretical descriptions of chemotactic drift, as there is a long history of modeling cell motion up shallow concentration gradients \cite{clark2005pnas,dufour2014ploscb,colin2014rsif,micali2017biophysj}. In the absence of noise, such mean-field approaches predict the scaling
\begin{equation}
    \frac{v_\text{drift}}{v} = \text{CI} = 
    \left\langle \cos \theta \right\rangle \sim g,
    \label{eq:v_drift_2}
\end{equation}
where CI is the chemotactic index, defined as the cosine similarity between the direction of motion and the gradient direction, and $\theta$ is the angle between them (Fig.~\ref{fig:fig1}A). Equation~\ref{eq:v_drift_2}, valid to linear order in $g$, is consistent with experimental data obtained in piecewise exponential gradients (Fig.~\ref{fig:fig2}A) \cite{micali2017biophysj}. For larger relative gradients, the drift increases, with distinct peaks corresponding to different major receptor types. Within each peak, the drift shows little dependence on absolute concentration, reflecting the log-sensing regime.

\color{black}

To understand the origin of the scaling in Eq. \ref{eq:v_drift_2}, the mean-field model is given by (see \cite{micali2017biophysj} for details)
\begin{equation}
v_\text{drift}(c,g) \approx C \left[
\nu_a \frac{c(K_a^{\mathrm{on}} - K_a^{\mathrm{off}})}{(c + K_a^{\mathrm{off}})(c + K_a^{\mathrm{on}})}
+
\nu_s \frac{c(K_s^{\mathrm{on}} - K_s^{\mathrm{off}})}{(c + K_s^{\mathrm{off}})(c + K_s^{\mathrm{on}})}
\right] g,
\label{eq:vd_approx}
\end{equation}
where $\nu_a$ and $\nu_s$ denote the fractions of Tar and Tsr receptors, respectively, with ligand dissociation constants $K_{a,s}^{\mathrm{on/off}}$ 
in the active and inactive states. Note, for $c>\!\!>K_X^\mathrm{off}$ and $c<\!\!<K_X^\mathrm{on}$ 
with $X=\{a,s\}$ this expression becomes effectively independent of ligand concentration. The prefactor $C$ captures the effective susceptibility of the chemotaxis pathway, coarse-grains the internal signaling response, and is proportional to $v$. In the full theory, this prefactor depends on the number of receptor dimers in a cluster and the pathway susceptibility evaluated at the adapted activity. In the present approximation, $C$ is treated as an effective constant and can be fixed by matching the peak drift velocity in the Tar-dominated regime. There is no collapse by rescaling with $\sqrt{\text{SNR}}$ (Fig.~\ref{fig:fig2}B) or $\text{SNR}$ (not shown).

\color{black}

Although Eq.~\ref{eq:v_drift_2} does not address threshold detection—namely, the smallest gradients that cells can reliably sense—it is nevertheless consistent with Weber's law of sensory perception. Weber's law states that the minimal detectable change in stimulus scales with the background level. Postulating a minimal detectable gradient $g_\text{min} = \Delta c/(l c) \approx \text{const}$, with the run length $l$ as a characteristic length scale, immediately yields $\Delta c \sim c$. Experimental measurements indeed suggest that bacterial chemotaxis follows this relation \cite{tu2008pnas,clausznitzer2014ploscb}. For larger stimuli, Weber's law generalizes to fold-change detection, which also applies to bacterial chemotaxis \cite{tu2008pnas,kalinin2010jbact}. Taken together, these results provide strong support for the $v_{\text{drift}} \sim g$ scaling reported by Mattingly \textit{et al.} in the log-sensing regime, rather than a scaling with the signal-to-noise ratio. To assess whether this behavior reflects a remarkable robustness to noise, it is instructive to contrast bacterial chemotaxis with gradient sensing in eukaryotic cells.

\section*{Lessons from eukaryotic chemotaxis}

Slower and larger eukaryotic cells, such as the slime mold \textit{Dictyostelium discoideum} (or neutrophils), are thought to sense chemical gradients spatially across their cell body (Fig.~\ref{fig:fig1}B). While it remains unclear whether such cells operate at the physical limits of sensing, their chemotactic performance is nevertheless expected to be constrained by noise, for example through the signal-to-noise ratio (SNR).

\begin{figure}[t]
\centering
\includegraphics[scale=0.4]{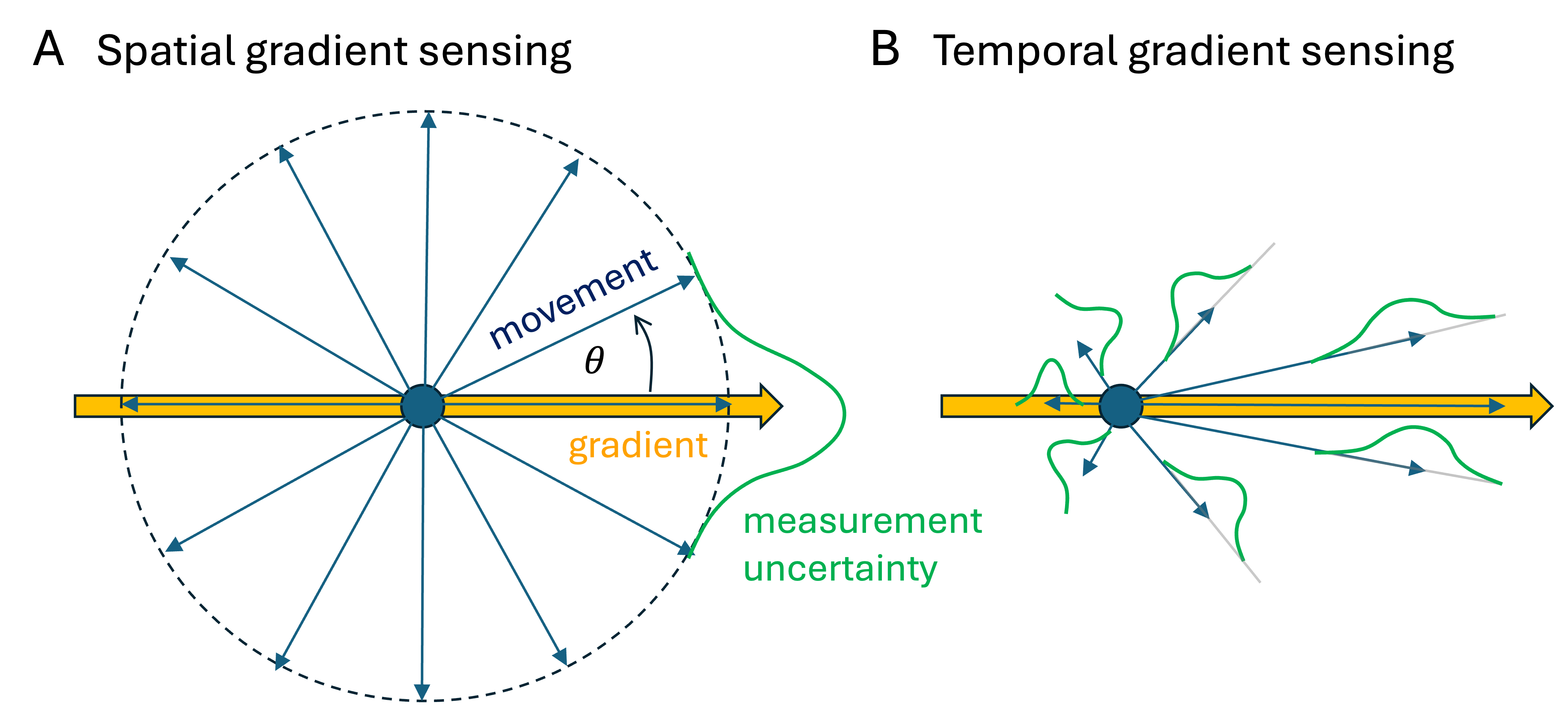}
\caption{{\bf Difference between spatial and temporal gradient sensing.} (A) In a minimal model for spatial gradient sensing in which direction of movement is biased (green distribution), not the "run length" (blue errors) \cite{endres2008pnas}. (B) In contrast, in a minimal model for temporal gradient sensing, the run length is biased but all directions are chosen by the cell with equal probability. Rotational diffusion is neglected here. The gradient points in the positive $x$ direction (yellow arrow).}
\label{fig:fig3}
\end{figure}

To make this connection explicit, one requires a model that translates physical sensing limits into measurable cell behavior. An analytically tractable model for the chemotactic index $\mathrm{CI}=\langle \cos\theta\rangle$ in shallow gradients was derived in Ref.~\cite{endres2008pnas}. In this framework, cells are assumed to estimate the spatial gradient by averaging positional information over a finite measurement time, before committing to movement in the inferred direction for a fixed distance (Fig.~\ref{fig:fig3}A). This procedure is repeated iteratively as the cell migrates up the gradient. For a perfectly ligand-absorbing cell, the probability distribution of gradient estimates is taken to be a bivariate Gaussian in two dimensions \cite{endres2008pnas}.

The resulting chemotactic index takes the saturating form
\begin{equation}
    \mathrm{CI} = \sqrt{\frac{\pi z}{2}}\, e^{-z}\,[ I_0(z) + I_1(z) ],
    \label{eq:optimal-similarity}
\end{equation}
where $z = 3\pi k\,\mathrm{SNR}$, with $\mathrm{SNR} = (\nabla c)^2/c$ (equivalently $g^2 c$), and $I_{0}$ and $I_{1}$ are modified Bessel functions of the zeroth and first order, respectively. The compound parameter $k = D a^3 T$ collects geometric and dynamical factors into a single fit parameter. When compared with experimental data, this expression predicts a collapse of the chemotactic index onto a single curve when plotted as a function of the SNR (Fig.~\ref{fig:fig2}C) \cite{endres2008pnas,alonso2025pnas}. Thus, in contrast to bacterial chemotaxis, eukaryotic chemotaxis appears to be directly constrained by the signal-to-noise ratio, consistent with additional experimental evidence \cite{amselem_control_2012}.

Motivated by this comparison, we can now return to bacterial chemotaxis and ask why no analogous SNR dependence emerges. A key difference is that bacteria do not attempt to estimate the direction of the gradient. Instead, in a minimal description of run-and-tumble chemotaxis, all run directions after a tumble are equally likely, and directional bias arises solely through modulation of run durations (Fig.~\ref{fig:fig3}B). Even when gradient estimation is noisy, the mean drift velocity remains
\begin{equation}
    \frac{v_\text{drift}}{v} = \frac{\alpha g}{2},
    \label{eq:final}
\end{equation}
where $\alpha$ depends on the run speed $v$ (see \textcolor{black}{Appendix A} for details). Importantly, to linear order in $g$ this result is unaffected by noise: fluctuations shorten some runs and lengthen others, but symmetry ensures that these effects cancel in the mean drift.

\textcolor{black}{The analysis of the scaling of eukaryotic chemotaxis can also be cast into the language of the information-theoretic toy model (see Appendix B for details). Briefly, in a spatial-sensing cell, the relevant behavioral output is not the modulation of run length but the estimation of movement direction. If the gradient estimate is noisy, the resulting angular uncertainty directly broadens the distribution of migration angles. Assuming for simplicity Gaussian directional errors with variance inversely proportional to the sensing SNR yields a chemotactic index $\mathrm{CI}=\langle \cos \theta \rangle$ that depends explicitly on SNR. Thus, unlike in the run-and-tumble case, no symmetry-based cancellation occurs, because noise acts directly on the directional decision.}

Consequently, unlike the chemotactic index used to describe eukaryotic chemotaxis \cite{endres2008pnas}, the bacterial drift derived here is remarkably insensitive to noise in shallow gradients. This highlights how run-and-tumble chemotaxis can function as a robust strategy in noisy environments, as previously suggested \cite{clark2005pnas,rode2024prxlife}. The present argument assumes linear response and Gaussian fluctuations; beyond this regime, correlated noise or strong gradients may introduce additional effects and alter behavior.

\section*{Conclusions and outlook}

Recent information-theoretic analyses suggest that bacterial chemotaxis operates far below the physical limits of sensing, with only a small fraction of available ligand information transmitted to downstream signaling \cite{mattingly2024natphys_submitted,mattingly2021natphys}. At face value, this result appears to challenge the long-standing view that bacterial navigation is near-optimal. Here, I have argued that this interpretation conflates information efficiency with behavioral performance. When viewed at the level of movement strategy, low information efficiency need not imply poor chemotactic ability.

%\textcolor{black}{To illustrate quantitatively, even though the diffusion-limited capture of ligand molecules can be large based on flux $J \sim 4\pi D a c \sim r_0$ \cite{berg1977biophysj}, implying relatively small fluctuations in absolute concentration measurements, the signal-to-noise ratio behaves as $\mathrm{SNR} \sim (g v \tau)^2 r_0 \tau$, which can remain small in shallow gradients even when the ligand flux $r_0$ is large. The present results show that, in this regime, the run-and-tumble strategy can nevertheless render chemotactic drift robust to such fluctuations.}

\textcolor{black}{To illustrate this point quantitatively, even though the diffusion-limited capture of ligand molecules can be large based on flux $J \sim 4\pi D a c \sim r_0$ \cite{berg1977biophysj}, implying relatively small fluctuations in absolute concentration measurements, the signal-to-noise ratio behaves as $\mathrm{SNR} \sim (g v \tau)^2 r_0 \tau$, which can remain small in shallow gradients even when the ligand flux $r_0$ is large. The present results show that, in this regime, the run-and-tumble strategy can nevertheless render chemotactic drift robust to such fluctuations.}

Using minimal models and scaling arguments, I showed how the run-and-tumble strategy can confer robustness to both external and internal noise through symmetry and temporal averaging. In shallow gradients, fluctuations in gradient estimation—at least for independent, additive noise—primarily redistribute run lengths rather than directions, leaving the mean drift largely unaffected. From this perspective, information loss along the signaling pathway may be a tolerable, or even inevitable, consequence of a strategy optimized for robustness rather than maximal information transmission. This interpretation is consistent with Weber’s law and with earlier mean-field models that successfully capture observed chemotactic drift \cite{tu2008pnas,clausznitzer2014ploscb,micali2017biophysj}.

\textcolor{black}{These considerations suggest a qualitative distinction between sensing strategies in different noise regimes. Temporal gradient sensing, as employed by bacteria, may be particularly advantageous in low-SNR environments, where averaging over time suppresses fluctuations and renders drift robust to noise. In contrast, spatial gradient sensing in eukaryotic cells allows for more direct responses at higher SNR, where directional estimates become reliable, but is more sensitive to noise at low SNR. This trade-off may provide a physical basis for the emergence of distinct chemotactic strategies across biological scales.}

Based on the above, future work may want to focus less on absolute information efficiency and more on how sensing, signaling, and movement strategies jointly determine chemotactic performance. Experimentally, direct measurements of drift scaling with gradient steepness, background concentration, and systematically varied sources of external (e.g. ligand arrival) and internal (signaling) noise would help disentangle these effects \cite{colin2014rsif}. In parallel, independent simulations of swimming bacteria with systematically varied noise sources could provide a complementary route to assessing how different forms of noise impact chemotactic behavior \cite{micali2017biophysj}. More broadly, contrasting temporal sensing in bacteria with spatial gradient sensing in slower and larger eukaryotic cells may help clarify how different strategies convert information into behavior across biological contexts \cite{WanJekely2021PRSB}.

Finally, while Mattingly \textit{et al.} compared their results with simulations of idealized cells that convert all available information into directed motion, such comparisons remain grounded in the same information-theoretic framework. By contrast, independent experimental analyses have found that chemotactic drift varies only weakly across wide ranges of environmental noise \cite{Brumley2019PNAS}, lending further support to the idea that robustness, rather than optimal information transmission, is the primary design principle underlying bacterial chemotaxis.

Taken together, these arguments highlight the importance of interpreting physical sensing limits in the context of the behavioral strategies that ultimately convert information into motion.

\color{black}

\section*{Appendix A: Mathematical derivation of mean drift velocity of bacteria with noise}

Here I calculate the mean drift velocity of run-and-tumble chemotactic bacteria in the presence of additive noise, thus deriving Eq.~\ref{eq:final}. I assume bacteria such as \textit{E. coli} swim with speed $v$ in a shallow relative spatial gradient $g$, and let $\theta$ denote the angle between the run direction and the gradient direction. Along a run, the spatial gradient is perceived as a temporal relative gradient $\dot c/c = v g \cos\theta$. Except for the inclusion of noise, the calculation follows standard approaches \cite{deGennes2004Chemotaxis,Schnitzer1993ContinuumRandomWalks,CelaniVergassola2010BacterialStrategies,micali2017biophysj}.

Run durations are stochastic, with distribution
\begin{equation}
P(t|\theta) = \lambda(t|\theta)\, P_\text{surv}(t|\theta)
= \lambda(t|\theta)\, \exp\!\left[-\int_0^t dt'\, \lambda(t'|\theta)\right],
\end{equation}
where $\lambda(t|\theta)$ is the (possibly history-dependent) instantaneous tumbling rate during a run at angle $\theta$ relative to the gradient. The mean run time is
\begin{eqnarray}
\langle \tau_r(\theta)\rangle &=& \int_0^\infty dt\, t\, P(t|\theta) \\
&\approx& \frac{1}{\lambda(\theta)}
= \frac{1}{\lambda_0\!\left(1-\alpha g\cos\theta\right)}
\approx \tau_0\!\left(1+\alpha g\cos\theta\right),
\end{eqnarray}
to linear order in $g$ for shallow gradients. In this regime the survival probability is effectively Markovian, runs are approximately exponentially distributed, and $\lambda_0=\tau_0^{-1}$ is the baseline tumbling rate in the absence of a gradient. The coefficient $\alpha$ is a proportionality constant that depends on physiological parameters, including run speed $v$.

The mean drift velocity satisfies $\langle v_\text{drift}\rangle = v\,\mathrm{CI}$, where $\mathrm{CI}=\langle\cos\theta\rangle$ is the chemotactic index (note in the main text we use $v_\text{drift}$ for simplicity of notation). Averaging over run directions weighted by the mean run time yields
\begin{equation}
\langle v_\text{drift}\rangle
= v\,\frac{\int_0^{2\pi} d\theta\, \langle\tau_r(\theta)\rangle \cos\theta}
{\int_0^{2\pi} d\theta\, \langle\tau_r(\theta)\rangle + \int_0^{2\pi} d\theta\, \langle\tau_t(\theta)\rangle}
\simeq v\,\frac{\alpha g \int_0^{2\pi} d\theta\, \cos^2\theta}{\int_0^{2\pi} d\theta}
= \frac{v \alpha g}{2} \propto g,
\end{equation}
where tumble times were neglected in the denominator because they are short compared with run times. This reproduces Eq.~\ref{eq:v_drift_2}. If rotational diffusion with rate $D_r$ is included, orientation decorrelation introduces an attenuation factor $1/(1+2D_r\tau_0)$.

Now suppose the perceived gradient, and hence the tumbling modulation, is noisy due to errors in temporal gradient estimation \cite{mora2010prl}. I model this by writing the mean run time conditional on a noisy gradient estimate as
\begin{equation}
\langle \tau_r(\theta)\rangle_\eta=\tau_0\left[1+\alpha\,(g+\eta)\cos\theta + \mathcal{O}(g^2)\right],
\end{equation}
where $\eta$ is additive Gaussian noise with $\langle \eta\rangle=0$ and variance $\langle \eta^2\rangle=\sigma^2$. Averaging over noise realizations then gives, to linear order in $g$, the (unconditional) mean drift 
\begin{equation}
\langle v_\text{drift}\rangle
= v\,
\frac{\int_0^{2\pi} d\theta\, \cos\theta \int \frac{d\eta}{\sqrt{2\pi\sigma^2}}
\,e^{-\eta^2/(2\sigma^2)}\left[1+\alpha(g+\eta)\cos\theta\right]}
{\int_0^{2\pi} d\theta \int \frac{d\eta}{\sqrt{2\pi\sigma^2}}
\,e^{-\eta^2/(2\sigma^2)}\left[1+\alpha(g+\eta)\cos\theta\right]}
= \frac{v \alpha g}{2} \propto g,
\label{eq:v_drift_noise}
\end{equation}
using $\int d\eta\, \eta\, e^{-\eta^2/(2\sigma^2)}=0$. Thus, symmetric additive noise—whether from measurement errors or intrinsic fluctuations—does not alter the mean drift to linear order: some runs become too short and others too long, but these effects cancel in the average. This illustrates why run-and-tumble chemotaxis can be remarkably robust in shallow gradients.

Beyond linear order in $g$, or for correlated noise $\eta(\theta)$ (for example arising from slow, history-dependent adaptation), this cancellation need not hold. For instance, at second order the denominator contains terms of the form
$\int_0^{2\pi} d\theta \int \frac{d\eta}{\sqrt{2\pi\sigma^2}}
\,e^{-\eta^2/(2\sigma^2)}\left[1+\tfrac{1}{2}\alpha^2(g+\eta)^2\cos^2\theta\right]$,
which no longer vanish even for additive white noise.

\section*{Appendix B: Information-theoretic toy model for spatial gradient sensing}

The spatial sensing problem can be rationalized within an information-theoretic framework by considering how much information is required to estimate the direction of a chemical gradient with finite precision.

In contrast to temporal sensing in bacteria, where the relevant output is a drift velocity, spatial sensing involves estimating a direction, characterized by an angle $\theta$. The accuracy of this estimate can be quantified by the variance $\sigma_\theta^2$ of angular errors.

Assuming that the gradient vector is inferred from noisy measurements, the signal-to-noise ratio (SNR) associated with this estimate determines the precision of directional sensing. To leading order, one expects
\begin{equation}
\sigma_\theta^2 \sim \frac{1}{\mathrm{SNR}},
\end{equation}
reflecting the fact that larger signals or reduced noise improve angular accuracy.

From an information-theoretic perspective, specifying a direction with uncertainty $\sigma_\theta$ requires a finite amount of information. For a Gaussian approximation of angular errors, this can be estimated as
\begin{equation}
I_{\mathrm{dir}} \sim -\frac{1}{2} \log(\sigma_\theta^2),
\end{equation}
up to an additive constant. Thus, higher information corresponds to smaller angular uncertainty.

The behavioral output is captured by the chemotactic index, $\mathrm{CI} = \langle \cos \theta \rangle$. For Gaussian angular fluctuations, one obtains
\begin{equation}
\mathrm{CI} \approx e^{-\sigma_\theta^2/2}.
\end{equation}
Combining this with the scaling of $\sigma_\theta^2$ yields
\begin{equation}
\mathrm{CI} \sim \exp\!\left(-\frac{\mathrm{const}}{\mathrm{SNR}}\right),
\end{equation}
showing that chemotactic performance depends explicitly on the signal-to-noise ratio.

This contrasts with temporal gradient sensing in run-and-tumble bacteria, where noise affects the modulation of run durations and can cancel by symmetry to leading order. In spatial sensing, by contrast, noise directly perturbs the estimated direction, leading to an intrinsic dependence of performance on SNR.

\color{black}

\section*{Acknowledgements}

I would like to mention previous funding from the UKRI Transition Award BB/W013770/1.

\bibliography{references}

\end{document}